\documentclass[11pt]{article}

\usepackage{authblk}
\usepackage{graphicx}
\usepackage{pdfsync}
\usepackage{hyperref}
\usepackage{statex2}

\begin{document}
%\runninghead{Logan et al.}
\title{Decision making and uncertainty quantification for individualized treatments}

\author[1]{Brent R. Logan}
\author[1]{Rodney Sparapani}
\author[2]{Robert E. McCulloch}
\author[1]{Purushottam W. Laud}
\affil[1]{Division of Biostatistics, Medical College of Wisconsin}
\affil[2]{School of Mathematical and Statistical Sciences, Arizona State University}
\date{}                     %% if you don't need date to appear
\setcounter{Maxaffil}{0}
\renewcommand\Affilfont{\itshape\small}

\maketitle
%\author{Brent R. Logan, 
%Rodney Sparapani, Robert E. McCulloch, and 
% Purushottam W. Laud}
%\affiliation{\affilnum{1} Division of Biostatistics, Medical College of Wisconsin \\
%\affilnum{2} School of Mathematical and Statistical Sciences, Arizona State University}
%\corrauth{Brent R. Logan, Division of Biostatistics, Medical College of Wisconsin\\
%8701 Watertown Plank Rd.\\
%PO Box 26509
%Milwaukee, WI 53226-0509}
%\email{blogan@mcw.edu}

\begin{abstract}
Individualized treatment rules (ITR) can improve health 
outcomes by recognizing that patients may respond 
differently to treatment and assigning therapy with the 
most desirable predicted outcome for each individual.  Flexible 
and efficient prediction models are desired as a basis for 
such ITRs to handle potentially complex interactions between 
patient factors and treatment. Modern Bayesian semiparametric and
nonparametric regression models provide an attractive avenue in this
regard as these allow natural posterior uncertainty quantification of
patient specific treatment decisions as well as the population wide
value of the prediction-based ITR. In addition, via the use of such
models, inference is also available for the value of the Optimal ITR. We propose
such an approach and implement it using  Bayesian Additive Regression 
Trees (BART) as this model has been shown to perform well in fitting 
nonparametric regression 
functions to continuous and binary responses, even with many 
covariates. It is also computationally efficient for use in
practice. With BART we investigate a treatment strategy which 
utilizes individualized predictions of patient outcomes from BART 
models.  Posterior distributions of patient outcomes under each 
treatment are used to assign the treatment that maximizes the expected 
posterior utility.  We also describe how to approximate such a
treatment policy with a clinically interpretable ITR, and 
quantify its expected outcome. The proposed method performs very well 
in extensive simulation studies in comparison with several 
existing methods. We illustrate the usage of the proposed 
method to identify an individualized choice of conditioning regimen 
for patients undergoing hematopoietic cell transplantation and
quantify the value of this method of choice in relation to the Optimal
ITR as well as non-individualized treatment strategies.   
\end{abstract}

%\keywords{Individualized treatment rules, prediction models, Bayesian Additive Regression  Trees (BART), boosting, random forests, outcome weighted learning, subgroup analysis, optimal ITR,value function estimation}

\maketitle

\section{Introduction}
There is increasing recognition in clinical trials that patients are
heterogeneous and may respond differently to treatment.  A major goal
of precision medicine is to identify which patients respond best to
which treatments and tailor the treatment strategy to the individual
patient.  This personalization of treatment based on patient clinical
features, biomarkers, and genetic information is formalized as an
individualized treatment rule (ITR) by Qian and Murphy \cite{QianMurp11}.
Individualized treatment rules extend classical subgroup analysis, in
which pre-specified subgroups of the population are assessed for
differential treatment effects, to the point where the treatment
benefit for each individual is used to determine a treatment
assignment rule that is, in some sense, optimal.  

Several strategies for identifying treatment rules have been proposed
in the literature.  Many of these utilize a model for the conditional
mean function of the outcome given treatment and covariates, and
optimize it over available treatments to define an ITR. Qian and
Murphy \cite{QianMurp11} show that good prediction accuracy of the 
outcome model is sufficient 
in order to ensure good performance of the associated ITR.  As
a result a number of strategies for flexible prediction models have
been proposed, including a large linear aproximation space with
penalization to avoid overfitting (\cite{QianMurp11,ImaiRatk13b}), generalized additive models (\cite{MoodDean14}),
boosting (\cite{KangJane14}), random forests (\cite{FostTayl11}), support vector regression (\cite{ZhaoZeng11}), kernel ridge regression (\cite{ZhanLabe15}),
and tree based methods or recursive partitioning
(\cite{DussConv10,DoovDuss14,LipkDmit11,SuTsai09,ZeilHoth08,LabeZhao15}).  An
alternative strategy is to directly optimize an estimator of the
expected outcome of a treatment rule over a class of potential rules
(\cite{XuYu15,ZhanTsia12,ZhaoZeng12,FuZhou16}). This has the
advantage of not requiring an accurate prediction model, but does
require specification of the class of treatment rules allowed. While
each method uses its own ``optimal'' choice, in this article we reserve
the phrase ``Optimal ITR'' for the ITR defined by Qian and Murphy
\cite{QianMurp11} in their Equation 1, and also stated in our Equation~\ref{eqn:optitr} below. 

The Bayesian framework leads to natural quantification
of uncertainty that allows construction of credible and prediction
intervals. A major distinguishing feature as described later is that 
our method provides direct inference on the value of the Optimal ITR 
while it is not clear how this can be done with other existing
methods. The Bayesian nonparametric regression method we have
implemented here is Bayesian Additive Regression Trees (BART)
(\cite{ChipGeor10}) to model the conditional mean function of the
outcome and to inform identification of an ITR. BART has been shown
to be efficient and flexible with performance comparable to or better
than its non-Bayesian competitors such as boosting, lasso, MARS, neural
nets and random forests (\cite{ChipGeor10}). Furthermore, our simulations
described later in the article show that BART ITR's performance was
better than or comparable to other existing methods in this setting of
identifying an ITR. Because of its tree based structure, BART can
effectively address interactions among variables, which is important
in this context for identification of treatment interactions leading
to differential treatment recommendations. In addition, recent
modifications to BART have been proposed that maintain excellent
 out-of-sample predictive performance even when a large number of
additional irrelevant regressors are added (\cite{Line17}). 

We present our work in the following sequence. In the next section we  
describe the notation for individualized treatment rules.  Following that, we review the BART methodology briefly. The next section titled ``BART for ITRs and Value Function Estimation'' describes our
proposed BART based ITR and addresses estimation of its value function
as well as that of the Optimal ITR.  Subsequently we show excellent performance of the proposed BART ITR in a benchmark simulation comparison to existing methods for ITRs.  In the next section we conduct additional simulations to examine the operating characteristics of the BART prediction model and the estimation of the value function.  In the section titled ``Summarizing the BART ITR'' we discuss ways to approximate the BART ITR to get an interpretable clinical rule for treatment assignment, similar to identification of subgroups of patients who would benefit from assignment to particular treatments.  The following section illustrates the usage of the proposed method on a medical application in hematopoietic cell transplantation.  The article ends with a discussion of our contribution as well as of some planned 
future developments.

\section{Individualized Treatment Rules}

Let $Y$ be the outcome of interest, with higher values of $Y$ being more desirable.  We focus on a binary outcome in this article, where $Y=1$ indicates a favorable outcome and $Y=0$ its complement, but the proposed method could easily be used with a continuous outcome as well.  Let $\mathcal{A}$ be the treatment space with $\mathcal{A}=\{-1,1\}$ and let $X=(x_1,\ldots,x_p)$ be the vector of patient characteristics being used to personalize treatment, with population space $\Omega_X$ defined so that $p(A=1|X=x)$ is bounded away from 0 and 1, i.e. $\Omega_X=\{X:p(A=a|X=x) \in (0,1), \forall a \in \mathcal{A} \}$.  We assume observations represent a random sample $(Y,A,X)$ either from a randomized trial or observational study.  For an observational study we make the usual assumptions to allow causal inference that treatment assignment is strongly ignorable, i.e. treatment $A$ is independent of the potential outcomes $Y|A=1$ and $Y|A=0$ given X.  Note that BART models have been proposed for use in observational studies in \cite{Hill11}.

A treatment rule $g(X)$ is a mapping from the covariate space $\Omega_x$ to the treatment space $\mathcal{A}$ so that patients with covariate value $X$ are assigned to treatment $g(X)$.  Let $P$ be the distribution of $(Y,A,X)$ and $E(Y)$ be the expectation with respect to $P$.  Let $P^g$ be the distribution of $(Y,A,X)$ given that $A=g(X)$ and let $E^g(Y)$ be the expectation with respect to $P^g$.  The value function $V(g)$ of a treatment rule $g(X)$ is the expected outcome associated with that treatment rule, i.e., $V(g)=E^g(Y)$.  An Optimal ITR $g_0$ is a treatment rule that optimizes the value function, 
\[ g_0 \in \mbox{arg max}_{g \in G} V(g),\]
where $G$ is a collection of possible treatment rules.  Since the value function can be written as 
\[ V(g)=E[E(Y|X,A=g(X))],\]
Qian and Murphy \cite{QianMurp11} show that the Optimal ITR satisfies 
\begin{equation}
\label{eqn:optitr}
 g_0(X) \in \mbox{arg max}_{a \in \mathcal{A}} E(Y|X,A=a) \mbox{\ a.s.},
\end{equation}
so that one possible solution is to assign to each patient the treatment which has the higher conditional expectation given their covariate vector $X$.
In practice, this strategy requires modeling of the conditional mean function and use of the estimated conditional mean function in the above expression to determine treatment assignment.  Qian and Murphy \cite{QianMurp11} showed that if the prediction error of such a model is small, then the reduction in value of the associated ITR $g$ compared to the Optimal ITR $g_0$ is also small, pointing to the need for a flexible and accurate prediction model for the conditional mean function.  

\section{BART methodology }\label{Methodology}

As BART is based on an ensemble of regression tree models, we begin with a 
simple example of a
regression tree model. 
We then describe how BART uses an ensemble 
of regression tree models for a numeric outcome.
Finally, we describe how the BART model for a numeric outcome is augmented to
model a binary outcome.

\begin{figure}
\centering
\includegraphics[scale=1.]{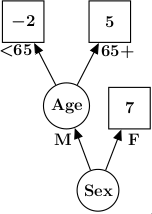}
\caption{An example of a single tree with branch decision rules and 
  terminal nodes\label{single-tree}}
\end{figure}

Suppose $y_i$ represents the (numeric) outcome for individual $i$, and $\bm{x}_i$ is a vector of covariates with the regression
relationship $y_i=h(\bm{x}_i; T, M)+\epsilon_i$.
Notationally, $h(\bm{x}_i; T, M)$ is a binary tree function with
components $T$ and $M$ that can be described as follows. 
$T$ denotes the tree structure consisting of two sets of nodes, interior
and terminal, and a branch decision rule at each interior node which typically
is a binary split based on a single component of the covariate vector.
An example is shown in Figure \ref{single-tree} wherein interior nodes appear
as circles, and terminal nodes as rectangles. The second tree component
$M=\{\mu_1, \dots, \mu_b \}$ is made up of the function values at the terminal 
nodes.

%\begin{figure}
%\centering
%\includegraphics[scale=0.4]{/home/blogan/bart/itr/ex4.png}
%\caption{An example of a single tree with branch decision rules and 
%  terminal nodes\label{single-tree}}
%\end{figure}

BART employs an ensemble of such trees in an additive fashion, i.e., it is the 
sum of $m$ trees where
$m$ is typically large such as 200. 
The model can be represented as:
\begin{equation}\label{eq:bartmodel}
\left. \begin{array}{rcl}
y_i        &=&f(\bm{x}_i)+\epsilon_i \where \epsilon_i \iid \N{0}{\sd^2} \\
f(\bm{x}_i)&=&\sum_{j=1}^m h(\bm{x}_i; T_j, M_j) 
\end{array}\right\}\ \ .
\end{equation}
To proceed with the Bayesian specification we need a prior for $f$. Notationally,
we use
\begin{equation}\label{eq:bartprior}
%f   \sim \BART \ 
f   \sim BART \ 
\end{equation}
and describe it as made up of two components: 
a prior on the complexity of each tree, $T_j$, and a
prior on its terminal nodes, $M_j|T_j$.  
Using the Smith-Gelfand bracket notation \cite{GelfSmit90} for distributions, 
we write $\wrap{f} = \prod_j \wrap{T_j} \wrap{M_j|T_j} $\ . Following
\cite{ChipGeor10}, we partition $\wrap{T_j}$ into 3 components: the
tree structure or process by which we build a tree and create interior nodes, the choice of a covariate given
an interior node and the choice of decision rule given a covariate for
an interior node.  
%The probability that a node at depth $d$ is
%interior is defined to be $\alpha (1+d)^{-\gamma}$ where $\alpha \in
%(0, 1)$ and $\gamma \ge 0$\ .  
The probability of a node being interior is defined by describing the 
probabilistic process by
which a tree is grown.
We start with a tree which is a single node and then recursively 
let a node have children (so that it is not a terminal node)
with probability 
$\alpha (1+d)^{-\gamma}$ where $d$ represents the branch depth, $\alpha \in
(0, 1)$ and $\gamma \ge 0$\ .  
We assume that the choice of a
covariate given an interior node and the choice of decision rule
branching value given a covariate for an interior node are both
uniform.  We then use the prior 
$\wrap{M_j|
T_j} = \prod_{\ell=1}^{b_j} \wrap{\mu_{j\ell}}$ 
where $b_j$ is the number of terminal nodes for tree $j$ and
%$\mu_{jk} \sim
%\N{0}{2.25/m}$ 
$\mu_{j\ell} \sim\N{0}{\tau^2/m}$ 
on the values of the terminal nodes. 
This gives $f(\bm{x}) \sim N(0,\tau^2)$ for any $\bm{x}$ 
since the value $f(\bm{x})$ will be the sum of $m$ independent
$N(0,\tau^2/m)$.
Along with centering
of the outcome, these
default prior mean and variance are specified such that each tree is a
``weak learner'' playing only a small part in the ensemble; more details
on this can be found in \cite{ChipGeor10}. 

To apply the BART model to a binary outcome, we use a probit transformation
$$
p(Y=1 \vert \bm{x}) \equiv p(\bm{x})=\Phi(\mu_0+f(\bm{x}))
$$
where $\Phi$ is the standard normal cumulative distribution function
and $f \sim BART$.
To estimate this model we use the approach of
Albert and Chib \cite{AlbeChib93} and augment the model with latent variables $Z_i$:
\begin{equation}\label{disp:binarybartmodel}
\begin{array}{rcl}
Y_i & = & I_{Z_i \ge 0} \\
Z_i & = & \mu_0 + f(\bm{x}_i) + \epsilon_i \\
f(\bm{x}_i)&=&\sum_{j=1}^m h(\bm{x}_i; T_j, M_j) \\
f & \sim & BART
\end{array}
\end{equation}
where $I_{Z \ge 0}$ is one if $Z \ge 0$ and zero otherwise 
and $\epsilon_i \sim N(0,1)$.
The Albert and Chib method then gives inference for $f$ using the Gibbs sampler
which draws $Z \vert f$ and $f \vert Z$.

%$p_i=E(Y_i)$, so that $f(x_i)=\Phi^{-1}(p_i)$.  
%A full specification is described below; note that the latent variables have unit variance so that there is no need to specify a prior for the variance.    
%\begin{equation}\label{disp:binarybartmodel}
%\left. \begin{array}{rcl}
%y_{i}|p_{i} & ~ & Bernoulli(p_{i}) \\
%p_{i}|f & = & \Phi(\mu_{i}),\ \mu_{i} =\mu_0+f(\bm{x}_i)  \\
%%f & ~ & \BART \\
%f & ~ & BART \\
%z_{i}|y_{i},f & ~ & \begin{cases} 
%\N{ \mu_{i}}{1}\I{-\infty, 0} & \If y_{i}=0 \\
%\N{ \mu_{i}}{1}\I{0, \infty} & \If y_{i}=1 \\
%\end{cases}  \\
%\end{array}
%\right\}\ \ .
%\end{equation}
The model just described can be readily estimated using existing software for
binary BART. It allows one to estimate the functions $f(\bm{x})$ 
through Markov Chain Monte Carlo draws of $f$
from which draws of the
corresponding success probabilities
$p(\bm{x})=\Phi(\mu_0+f(\bm{x}))$ 
are readily obtained.
Here $\mu_0$ is a tuning parameter to be chosen.
For example, setting $\mu_0=0$ centers the prior for $p(\bm{x})$ at .5 since the BART prior for $f$ is centered at 0; however, the BART model is fairly robust to different prior values of $\mu_0$ given sufficient data.
In the binary probit case, we let $\tau=1.5$. With this choice there is a 95\% prior probability that $f(\bm{x})$
is in the interval $\pm 2(1.5)$ giving a reasonable range of values for $p(\bm{x})$.

\section{BART for ITRs and value function estimation}

Due to the excellent flexibility in modeling complex interactions and 
the strong predictive performance, 
we propose to use individualized predictions of patient outcomes from the BART model 
to determine an ITR.  
In addition to constructing an ITR based on BART predictions,
we use the BART MCMC to assess uncertainty about various aspects
of the treatment decision and resulting value function.

In applying BART to the ITR problem,
our ``$\boldmath{x}$'' of the previous section 
(in which we reviewed BART)
becomes $(x,a)$ where, as in the section introducing ITR's, 
$a$ is the decision
variable and $x$ is patient information.

The BART model implies
$p(Y=1 \vert x,a,f) = \Phi(\mu_0 + f(x,a))$
where $f$ is expressed as the sum of trees as in Equation (\ref{disp:binarybartmodel}).
%The parameter of our model is the function $f$.
Posterior MCMC draws  $\{T^d_j,M^d_j\}$ consist of the individual
trees $j=1,2,\ldots,m$, for MCMC iterations $d=1,2,\ldots,D$.
Using Equation (\ref{disp:binarybartmodel}), each of these draws results in a
draw $f_d$ of the function $f$. In this section, it is helpful to view
the function $f$ as the fundamental underlying parameter.  BART is
viewed as giving us draws $\{f_d\}_{d=1}^D$ from the posterior 
distribution of $f$.

\subsection{Decision Theoretic Predictive BART ITR}
From Equation (\ref{eqn:optitr}), the Optimal ITR is obtained by choosing
the value of $a$ which maximizes
$E(Y \vert x,a) = p(Y=1 \vert x,a)$ conditional on $X=x$.
In our Bayesian framework, $p(Y=1 \vert x,a)$ is the predictive probability that $Y=1$
which is obtained by integrating $p(Y=1 \vert x,a,f)$ over the posterior distribution
of the parameter $f$.  Given MCMC draws $\{f_d\}$ from the posterior distribution of $f$, this integral is approximated by
averaging over the draws:
\begin{equation}
\label{eqn:predprob}
p(Y=1 \vert x,a) 
 \approx  \frac{1}{D} \sum_{d=1}^D \, p(Y=1 \vert x,a,f_d) 
 \equiv  \bar{p}(x,a).
\end{equation}
Thus, $\bar{p}(x,a)$ is the MCMC estimate of the predictive $p(Y=1 \vert x,a)$.  Illustrations of inference on patient specific predictive probabilities are shown later in the Example Section and in Figure~\ref{fig:waterfall}.

We can now define the BART based ITR as the one in which the treatment for each individual 
is given by maximizing 
the patient specific predictive probability
over available treatments:
%is based on the maximum posterior mean, 
\begin{equation}
\label{eqn:BARTITR}
g_{\mbox{\tiny{BART}}}(x)=\arg \max_a \bar{p}(x,a).
\end{equation}

The construction of $g_{\mbox{\tiny{BART}}}$ follows the basic
prescription of Bayesian decision theory in which we pick
the action which maximizes expected utility.
In this case, our utility is the outcome $Y$, and because $Y$ is binary, the expected $Y$
is the probability that $Y=1$, so we simply pick the action which
gives us the highest predictive probability of a successful outcome.

\subsection{Posterior Distribution of the Value function of an ITR}

For any ITR $g(x)$ it is of interest to assess its value across the patient population so that different ITRs may be compared via this value.  With an underlying function $f$, the value of the ITT $g$ is defined as 
$$
V(g,f) = E_X(p(Y=1 \vert x,g(x),f),
$$
which is the average (over $x$) of the probability of a good outcome.  Given MCMC draws we can
approximate the marginal distribution of this function of the
uncertain $f$ by simply plugging in draws of $f$:
\begin{equation}
\label{eqn:valuefn}
V_d(g) = V(g,f_d). 
\end{equation}
The posterior samples $\{V_d(g)\}, \; d=1,2,\ldots,D$, provide
inference for the value function of any ITR $g$, including the BART ITR
in Equation (\ref{eqn:BARTITR}), approximating the expectation over
$X$ by an average over a representative distribution of $x$ which is
often taken to be the observed samples in the covariate space. Illustrations
of inference on the value function can be found later in the Example
Section and in Figure~\ref{fig:waterfall}.

\subsection{Posterior Distribution of the Value function of the Optimal ITR}

It is also possible to estimate and assess uncertainty of the value
associated with the optimal ITR as defined in Equation (\ref{eqn:optitr}).  
We consider the optimal ITR as a function of $f$.  
If we knew $f$ then, given $x$, the optimal action is given by
$$
a(x,f) = \arg \max_a p(Y=1 \vert x,a,f)
$$
with corresponding maximum success probability $p^*(x,f)=\max_a p(Y=1
\vert x,a,f)$. Draws of the value of the Optimal ITR, namely
$V^*=E_X(p^*(x,f))$ can be obtained from draws of $f$ as
$$
V^*_d = E_X( p^*(x,f_d) ),\ \ d=1,\dots,D.
$$
Again, the expectation over $X$ is typically an average of
representative $x$ values, often using the observed samples in the
covariate space. We can interpret the draws $V_d$ collectively as
representing uncertainty about the value of the Optimal ITR, i.e., the
ITR for an agent who acts knowing the true $f$. Illustration
of this is also found later in the Example Section and Figure~\ref{fig:waterfall} (c) and (d). 

\section{Comparison with Existing Methods}

We conducted extensive simulation studies benchmarking the proposed BART ITR strategy against existing methods for identifying ITRs.  In order to avoid any concerns that we are deliberately selecting simulation settings where BART will show good performance, we reproduced simulation settings from two recent papers (\cite{XuYu15,KangJane14}) with a binary outcome variable $Y$ and binary treatment $A$.  The first set of simulations from \cite{XuYu15} included 5 additional binary covariates $X_A:X_E$, 5 ordinal covariates $X_a:X_e$ with 4 categories each, and one or two continuous covariates $X_{Ca},X_{Cb}$.  Eight different scenarios for the logit of the probability of response were simulated according to the following: 
\begin{itemize}
\item[(A)] $0.5X_{C1}+2(X_{B1}+X_{a3}*X_{A1})*A$ 
\item[(B)] $0.5X_{C1}+2(X_{B1}+X_{a3}*(X_{b2}+X_{b3}))*A$ 
\item[(C)] $0.05(-X_{A1}+X_{B1})+[(X_{a2}+X_{a3})+(X_{b2}+X_{b3})*X_{Ca}]*A$
\item[(D)] $\log \log [(X_{b3}+X_{c3})+5(X_{a2}+X_{a3}+X_{A1}X_{B1})*A+20]^2$
\item[(E)] $(X_{A1}+X_{B1})+2*A$
\item[(F)] $0.5X_{A1}+0.5X_{B1}+2I(X_{Ca}<5,X_a<2)*A$
\item[(G)] $0.5X_{A1}+0.5X_{B1}+2I(X_{Ca}<5,X_{Cb}<2)*A$
\item[(H)] $0.5X_{Ca}+0.5X_{Cb}+2I(X_{Ca}<-2,X_{Cb}>2)*A$
\end{itemize}
We also considered a modification of these scenarios (denoted (A2-H2)) in which the treatment interaction term was reduced to 1/4th of the given value, in order to better differentiate among the competing methods.  
In the second set of simulations from \cite{KangJane14}, up to 3 independent continuous markers $X_1,X_2,X_3$ were included in seven different models for the probability of nonresponse according to the following:
\begin{itemize}
\item[(K1)] $\logit p(Y=0|A,X)=0.3+0.2X_1-0.2X_2-0.2X_3+A(-0.1-2X_1-0.7X_2-0.1X_3)$, $X_j \sim N(0,1)$ 
\item[(K2)] $\logit p(Y=0|A,X)=0.3+0.2X_1-0.2X_2-0.2X_3+A(-0.1-2X_1-0.7X_2-0.1X_3)$, $X_j \sim N(0,1)$ except for 2\% of high leverage points with $X_1 \sim Uniform (8,9)$.  
\item[(K3)] $\log(-\log p(Y=0|A,X))=-0.7-0.2X_1-0.2X_2+0.1X_3+A(0.1+2X_1-X_2-0.3X_3)$, $X_j \sim N(0,1)$
\item[(K4)] $\log(-\log p(Y=0|A,X))=2-1.5X_1^2-1.5X_2^2+3X_1X_2+A(-0.1-X_1+X_2)$, $X_j \sim Uniform(-1.5,1.5)$
\item[(K5)]  $\logit p(Y=0|A,X)=-0.1-0.2X_1+0.2X_2-0.1X_3+X_1^2+A(-0.5-2X_1-X_2-0.1X_3+2X_1^2)$, $X_j \sim N(0,1)$ 
\item[(K6)]  $\logit p(Y=0|A,X)=0.1-0.2X_1+0.2X_2-X_1X_2+A(-0.5-X_1+X_2+3X_1X_2)$, $X_j \sim N(0,1)$ 
\item[(K7)] $p(Y=0|A,X)=I(X_1<8)(1+e^{-\eta})^{-1}+I(X_1 \geq 8)(1-(1+e^{-\eta})^{-1})$, where $\eta=0.3+0.2X_1-0.2X_2-0.2X_3+A(-0.1-2X_1-0.7X_2-0.1X_3)$ and $X_j \sim N(0,1)$ except for 2\% of high leverage points with $X_1 \sim Uniform (8,9)$. 
\end{itemize}
In all cases, ITRs were generated using a training dataset with $n=500$ observations, and then each ITR was applied to a fixed independent test dataset of 2000 observations in order to compute the value function for this ITR from the true model.  This process was replicated using 50 training datasets, and the average value function across the 50 training sets was obtained.  This average value function for a particular ITR was normalized as a fraction of the true optimal value function to facilitate comparisons across scenarios.   

For the BART ITR, we considered both use of the default prior parameters (BARTd), as well as cross-validation to select the number of trees (m=80,200) and a divisor for the prior SD $\tau$, denoted $k=(0.1,0.4,1.0)$ with default value of 1 (BARTcv).  Several competing methods were included for comparison, including regularized outcome weighted subgroup identification (ROWSI) (\cite{XuYu15}), Outcome Weighted Learning (OWL) (\cite{ZhaoZeng12}), use of random forests (RF) for outcome prediction along the lines of virtual twins approach (\cite{FostTayl11}) with cross validation of number of trees and minimum node size, and boosting with classification tree working model (KANG) (\cite{KangJane14}).  Ordinal variables were handled as ordinal for BART and other tree based methods, but were otherwise treated as categorical variables.    

\begin{figure}
\begin{center}
\includegraphics[height=15cm]{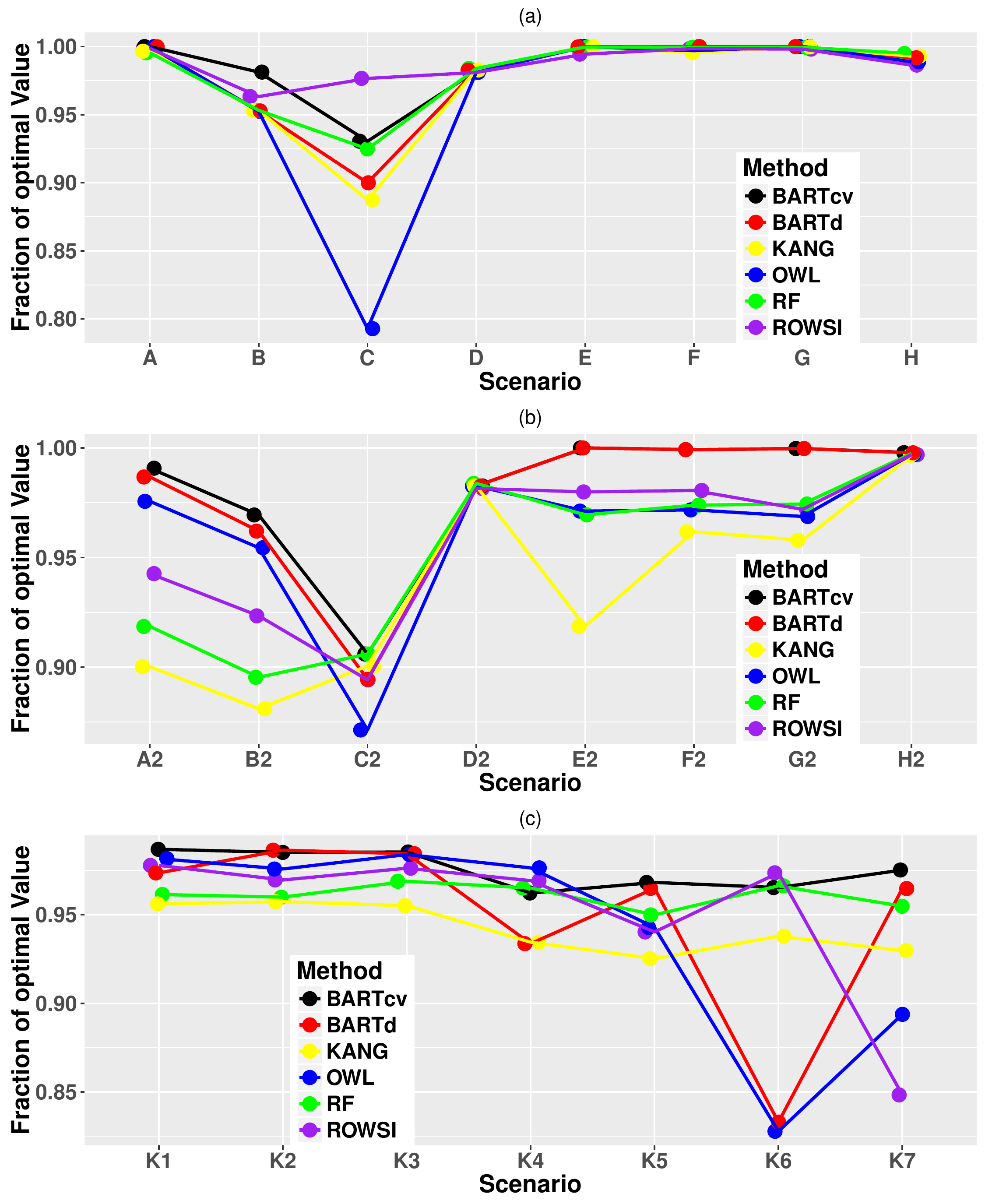}
\end{center}
\caption{Value function relative to the optimal value function for simulation settings in (a) scenarios as in \cite{XuYu15}, (b) same as (a) but with treatment effects cut by 25\%, and (c) scenarios as in \cite{KangJane14}}
\label{fig:simscenarios}
\end{figure}

Results are shown in Figure~\ref{fig:simscenarios}.  In all cases, the value function of the BART ITR with cross-validation performed at or near the top of the competing methods.  BART with the default settings also performed comparably to the other existing methods, with good performance in most situations.  

\section{Illustration of Operating Characteristics of BART prediction models and Estimation of Optimal Value Function}

\begin{figure}
\begin{center}
\includegraphics[height=15cm]{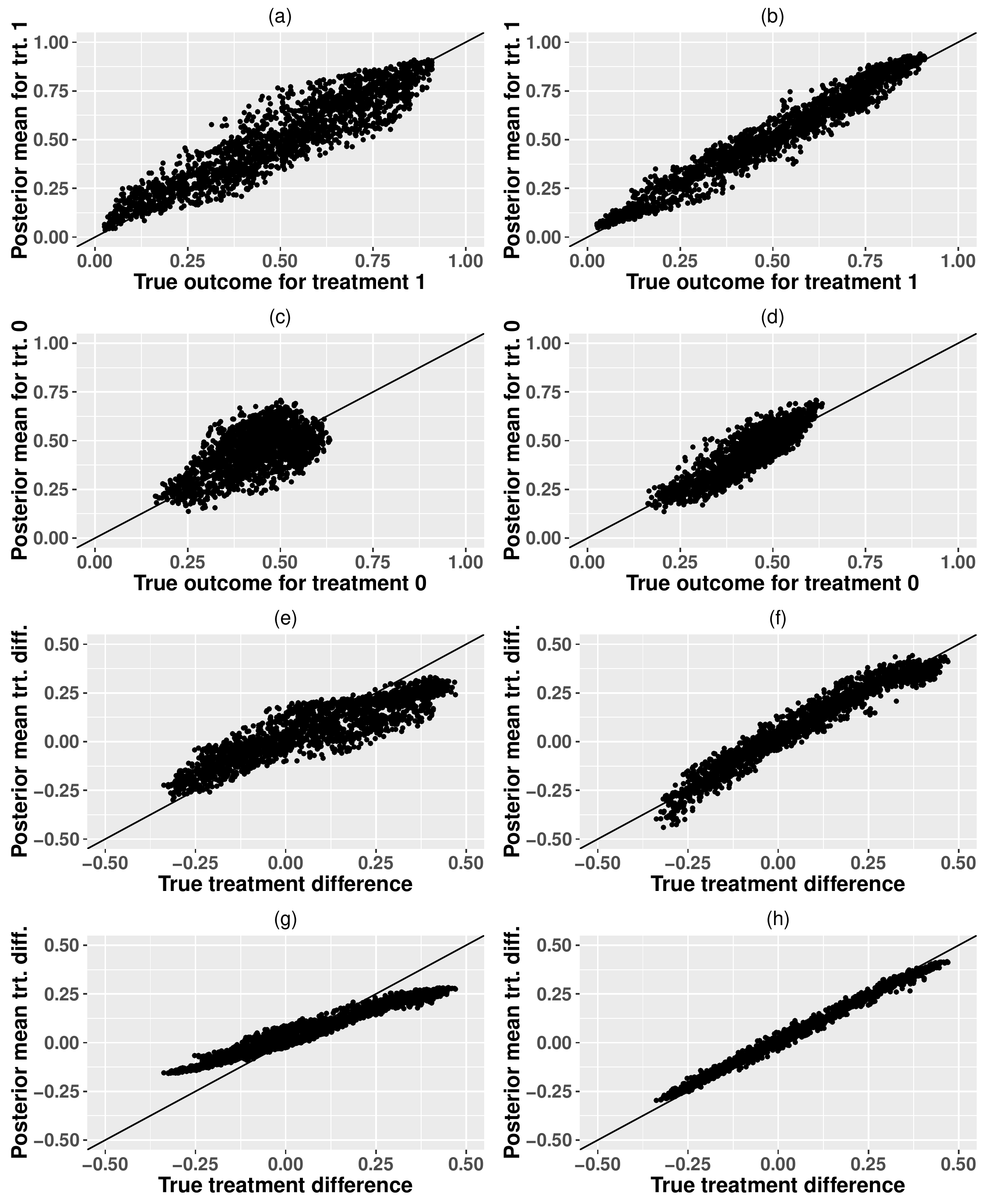}
\end{center}
\caption{BART posterior means vs. true probabilities for complex interaction model, with $n=500$ (left side) and $n=5000$ (right side).  First two rows show predictions for individual treatment outcomes, while row 3 shows predictions for treatment differences, all using a single training dataset.  Row 4 shows the posterior means for the treatment difference averaged over $400$ repeated data simulations of the training set.}
\label{fig:simdemo}
\end{figure}

We demonstrate the features of the proposed method using generated data from two settings: a complex treatment interaction setting, and a main effect only setting.  In each case, training datasets of either $n=500$ or $n=5000$ were generated and applied to an independent test dataset of $2000$ observations.  Logistic regression with a binary outcome and three independent Uniform$(-1.5,1.5)$ covariates was used to generate the data, with a complex treatment interaction model according to 
\begin{eqnarray*} \lefteqn{P(Y=1|A,X)=\left[1+\exp\left\{-0.1-0.2X_1+0.2X_2-0.1X_3+0.5X_1^2 \right. \right. } \\
&& \left. \left. +A(-0.5-0.5X_1-X_2-0.3I(X_3>0.5)+0.5X_1^2)\right\}\right]^{-1}.\end{eqnarray*}
and a no treatment interaction model according to 
\[  P(Y=1|A,X)=\left[1+\exp\{-0.1-0.2X_1+0.2X_2-0.1X_3+0.5X_1^2-0.3A\} \right]^{-1}.\]
In Figure~\ref{fig:simdemo} we plot the BART posterior means vs. the true probabilities of treatment outcome for the test dataset, for each treatment as well as for the treatment difference, for single training datasets of size $n=500$ ((a),(c),(e)) and $n=5000$ ((b),(d),(f)) using the complex interaction model.  The posterior mean prediction from the BART model has excellent accuracy for the large sample size, both for individual treatment outcomes as well as for the treatment difference, despite the complex treatment interaction model which includes linear and quadratic covariate-treatment interactions as well as an interaction term with a thresholded value of a covariate.  The larger training dataset shows improved accuracy compared to the smaller training dataset, which has some modest shrinkage of the treatment effects.  

We also conducted repeated data simulations with 400 replicates to
look at the bias of the prediction model as well as coverage of the
95\% interval estimates for the value function of the Optimal ITR,
using the quantiles of the posterior samples for the value function of
the Optimal ITR.  Bias over 400 replicates for the treatment
difference is shown in Figure~\ref{fig:simdemo}, panels (g) and (h).
As is seen with the single training dataset, there is some small bias
and shrinkage of the treatment effects for smaller sample size which
disappears with larger sample size.  Coverage probabilities of 95\%
credible intervals for the value of the Optimal ITR are 90\% for training dataset of size $n=500$ and 95\% for training dataset of size $n=5000$, indicating that once the sample size is sufficient to reduce the shrinkage of the treatment effects, coverage of the value function for the Optimal ITR is excellent. 
 
\begin{figure}
\begin{center}
\includegraphics[height=15cm]{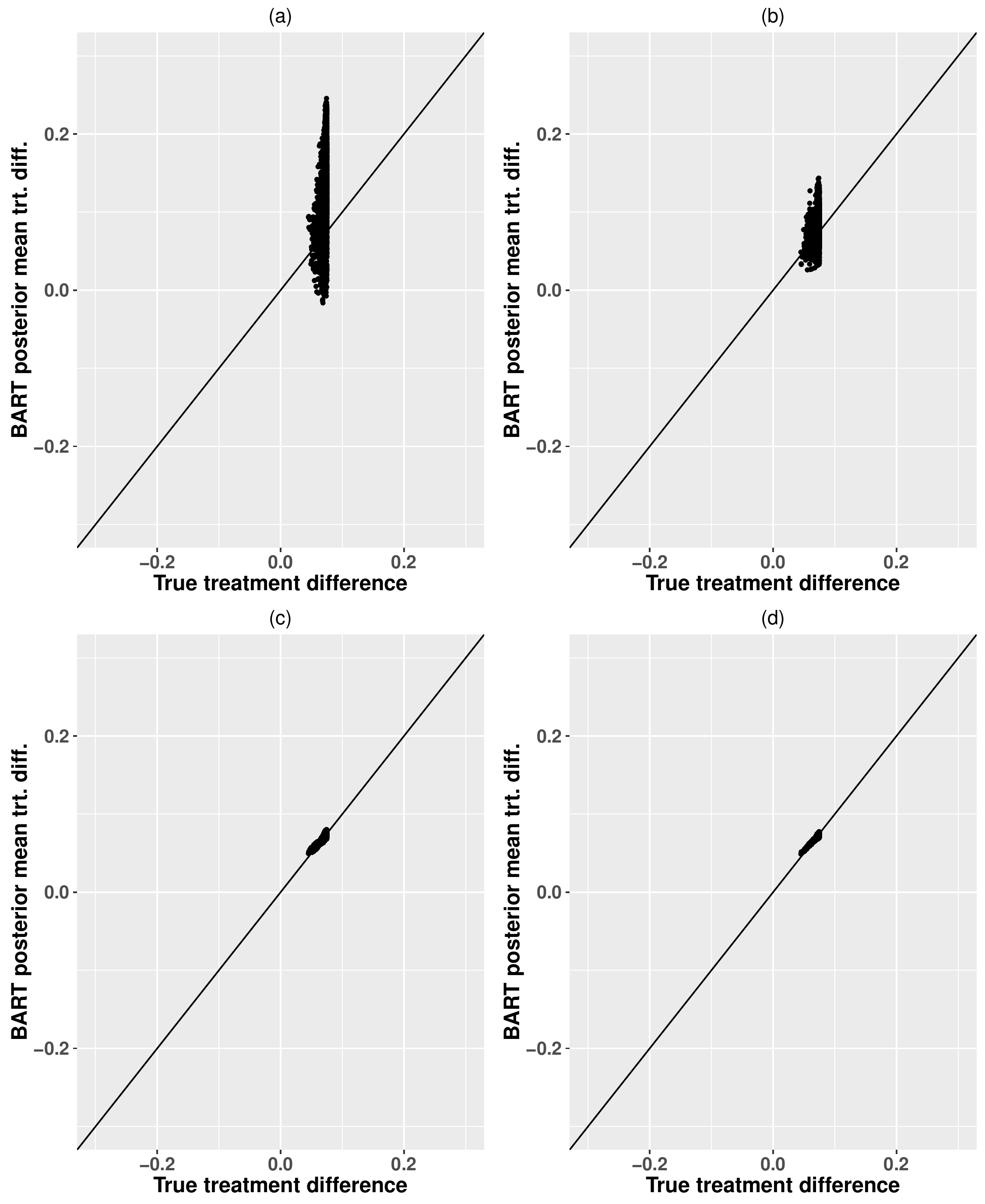}
\end{center}
\caption{BART posterior means vs. true probabilities for no interaction model, with $n=500$ (left side) and $n=5000$ (right side).  First row shows predictions for treatment differences using a single training dataset, while row 2 shows the posterior means for the treatment difference averaged over $400$ repeated data simulations of the training set.}
\label{fig:nullsimdemo}
\end{figure}

Similar results are shown for the model with no treatment-covariate interaction in Figure~\ref{fig:nullsimdemo}.  Note that the true treatment differences show very narrow variability due to the data generating model, and the BART model shows predictions for treatment differences which are also small and which narrow with increasing sample size.  These predictions are unbiased in repeated simulation, as indicated by the convergence to the diagonal line in Figure~\ref{fig:nullsimdemo} panels (c) and (d).  

\section{Summarizing the BART ITR}
The ITR based on the BART prediction model does not directly yield a
simple interpretable rule; this issue in general with flexible models has been 
discussed in \cite{ZhanLabe15}, who propose directly optimizing the value function over an 
interpretable set of rules.  In contrast, we separate the modeling of outcome from the determination of an
interpretable rule, by trying to develop an
approximation to this BART ITR which is interpretable and yields good
performance.  We propose a ``Fit-the-fit'' strategy, in which one
develops a single tree fit to the posterior mean treatment differences
as a function of patient characteristics.  Essentially, the posterior
mean treatment differences are treated as the ``data'', and we try and
fit an interpretable single tree to this data.   The single tree then
provides an interpretable way to explain which groups of patients
should receive which treatment, as well as the magnitude of the
treatment difference for that group of patients.  This strategy was
originally proposed as a variable selection technique for a BART
prediction model, but has been adapted here to focus on summarizing
the inference on treatment differences and the BART ITR.  

To fit an appropriate tree and also identify the best set of variables to include in that tree, a sequence of trees are fit, where variables are sequentially added to the candidate set of splitting variables in a stepwise manner to improve the fit.  Given a current set of variables and a corresponding tree built using those variables, the fit of the current tree is assessed using the $R^2$ between the fitted values from the tree and the posterior mean treatment difference (which is being used as the ``data'').  Additional variables are considered to be added one at a time yielding new trees built with an increasing set of variables, and their $R^2$ is assessed for each new variable added and corresponding tree.  The variable which most increases the $R^2$  is selected at each step.  Once
the $R^2$ does not improve appreciably with addition of a new variable
(we used $\Delta R^2<1\%$), the procedure ends and the current tree is
used as the approximation to the BART ITR. 

Note that each single tree
fit can be implemented very quickly, so this fit the fit
postprocessing procedure takes minimal computing time. It may not
always be possible to identify such a single tree (this is in fact the
benefit of ensemble methods to provide improved prediction), but note
that the quality of such an interpretable approximation can be
assessed using the $R^2$ between the single tree and the BART
prediction model. It is also possible to compute the value function of
the single-tree approximation ITR, and compare it with that of the
full BART-ITR.

\section{Example}

We illustrate the proposed methodology with a study of 1 year survival outcomes after a hematopoietic cell transplant (HCT) used to treat a variety of hematologic malignancies.    We use data on 3802 patients receiving reduced intensity conditioning for their HCT between 2011-2013, with data reported to the Center for International Blood and Marrow Transplant Research (CIBMTR).  Follow up to 1 year is complete for all patients so we analyze the outcomes using binary methods as described throughout the paper applied to the patient's survival status at 1 year.  Note that a full survival analysis could also be conducted with BART methods available for survival data(\cite{SparLoga16}), however this requires consideration of the appropriate target outcome to optimize and so we defer this for future research.  The primary treatment of interest is the type of conditioning regimen used (Fludarabine/Melphalan, or FluMel for short, vs. Fludarabine/Busulfan, or FluBu for short).  A variety of patient, donor, and disease factors were examined for their utility in personalizing the selection of the conditioning regimen, including age, race/ethnicity, performance score, Cytomegalovirus status, disease, remission status, disease subtypes, chemosensitivity, interval from diagnosis to transplant, donor type, Human Leukocyte Antigen (HLA) matching between donor and recipient, prior autologous transplant, gender matching between donor and recipient, comorbidity score, and year of transplant.  This observational cohort appears to be well balanced between the regimens across these factors, indicating reasonable equipoise by clinicians on which conditioning regimen is most appropriate for individual patients.  

\begin{figure}
\begin{center}
\includegraphics[height=15cm]{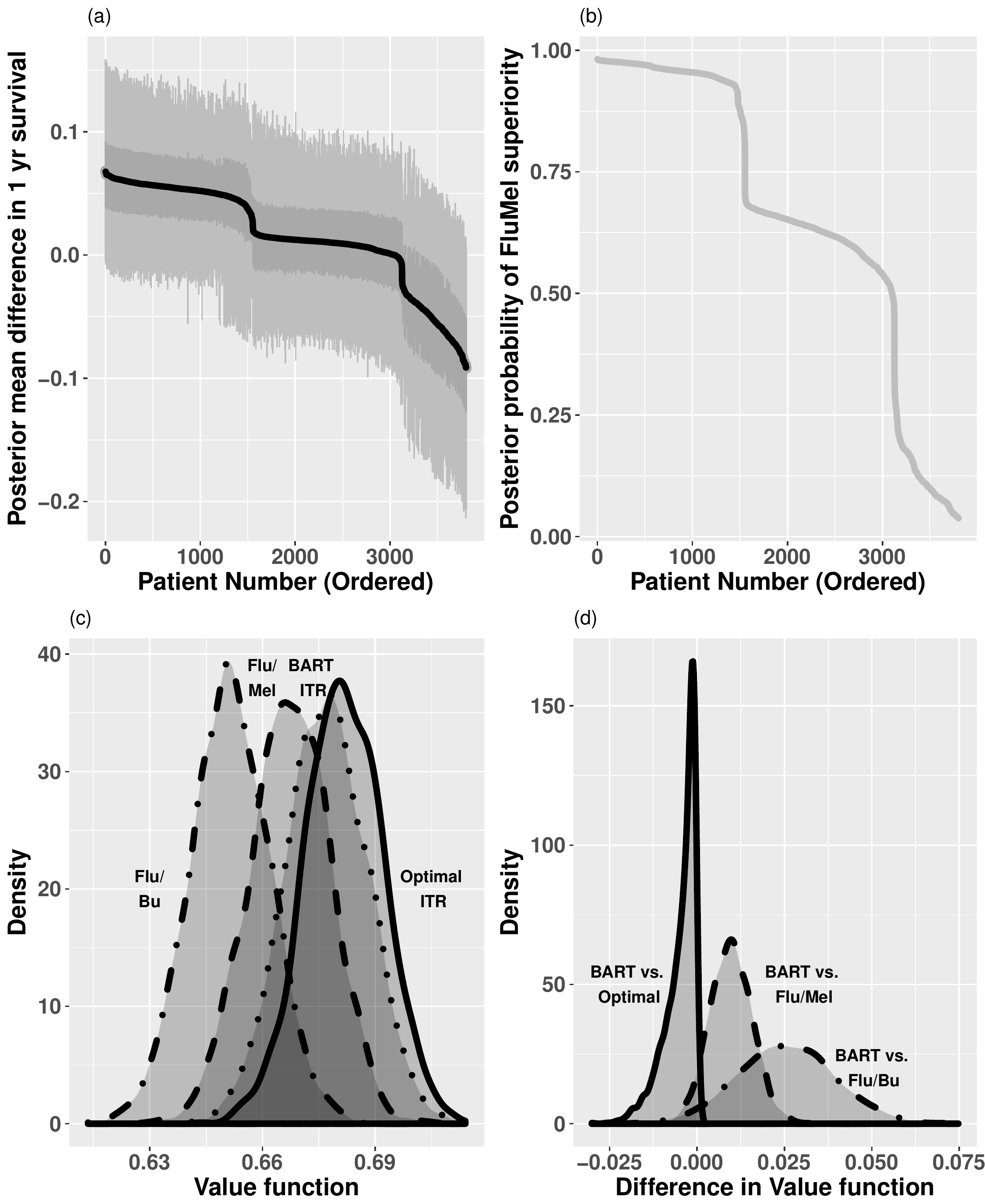}
\end{center}
\caption{Results of BART ITR for HCT example.  (a) Waterfall plot of 1 yr survival differences (FluMel-FluBu) by patient (posterior mean differences, along with inter-quartile ranges and 95\% posterior intervals), (b) Waterfall plot of posterior probabilities that survival is higher for Flu/Mel, (c) Density plot of value functions for three treatment strategies (FluMel, FluBu, BART ITR) as well as Optimal ITR, and (d) Density plot of difference in value functions for treatment strategies compared to BART ITR.  The posterior mean of the value function distributions for each treatment strategy are: FluBu: 0.651, FluMel: 0.667, BART ITR: 0.677, Optimal ITR: 0.682.  }
\label{fig:waterfall}
\end{figure}

Fitting of the BART model provides samples from $p(Y=1|x,a,f_d)$ for $d=1,\ldots,D$.  In Figure~\ref{fig:waterfall} we show waterfall plots of the differences in 1 year survival between Flu/Mel and Flu/Bu conditioning across patients in two ways.  In (a) we use the samples from the patient specific difference in 1 year survival, $p(Y=1|x,a=\mbox{Flu/Mel},f_d)-(Y=1|x,a=\mbox{Flu/Bu},f_d)$, and plot the posterior mean of these differences in 1 year survival for each patient, sorted by the magnitude of the difference.  This is equivalent to the difference in predictive probabilities under each treatment condition as described in Equation~\ref{eqn:predprob}.  Inter-quartile ranges and 95\% posterior intervals are also shown to indicate the variability of the differences. In (b) we show the waterfall plot of the posterior probabilities that Flu/Mel has a higher 1 year survival than Flu/Bu.  This is obtained by computing 
\[ \frac{1}{D} \sum_d I(P(Y=1|x,a=\mbox{Flu/Mel},f_d)>P(Y=1|x,a=\mbox{Flu/Bu},f_d)).\]
These plots indicate some heterogeneity in treatment benefit, where approximately 3000 of the patients seem to benefit from Flu/Mel, albeit with varying degrees of magnitude and/or certainty surrounding that benefit, while the remaining patients seem to benefit from Flu/Bu conditioning. 

Also in Figure~\ref{fig:waterfall} (c) we show the value functions as described in Equation~\ref{eqn:valuefn} for the cohort of $n=3802$ patients for three treatment rules: all patients receive Flu/Bu, all patients receive Flu/Mel, and patients receive treatment according to the BART based ITR.  The BART ITR value function distribution is shifted to the right, indicating improved 1 year survival outcomes over the overall cohort using this individualized strategy. The posterior mean of the value function distributions for each treatment strategy are: FluBu: 0.651, FluMel: 0.667, BART ITR: 0.677, Optimal ITR: 0.682. Figure~\ref{fig:waterfall} (d) shows the density functions of the difference in value function for the BART ITR compared to the other strategies, indicating high likelihood that the BART ITR is superior to the fixed treatment strategies.  On the other hand, it is inferior to the Optimal ITR as expected,  but the differences are small.

\begin{figure}
\begin{center}
\includegraphics[width=11cm]{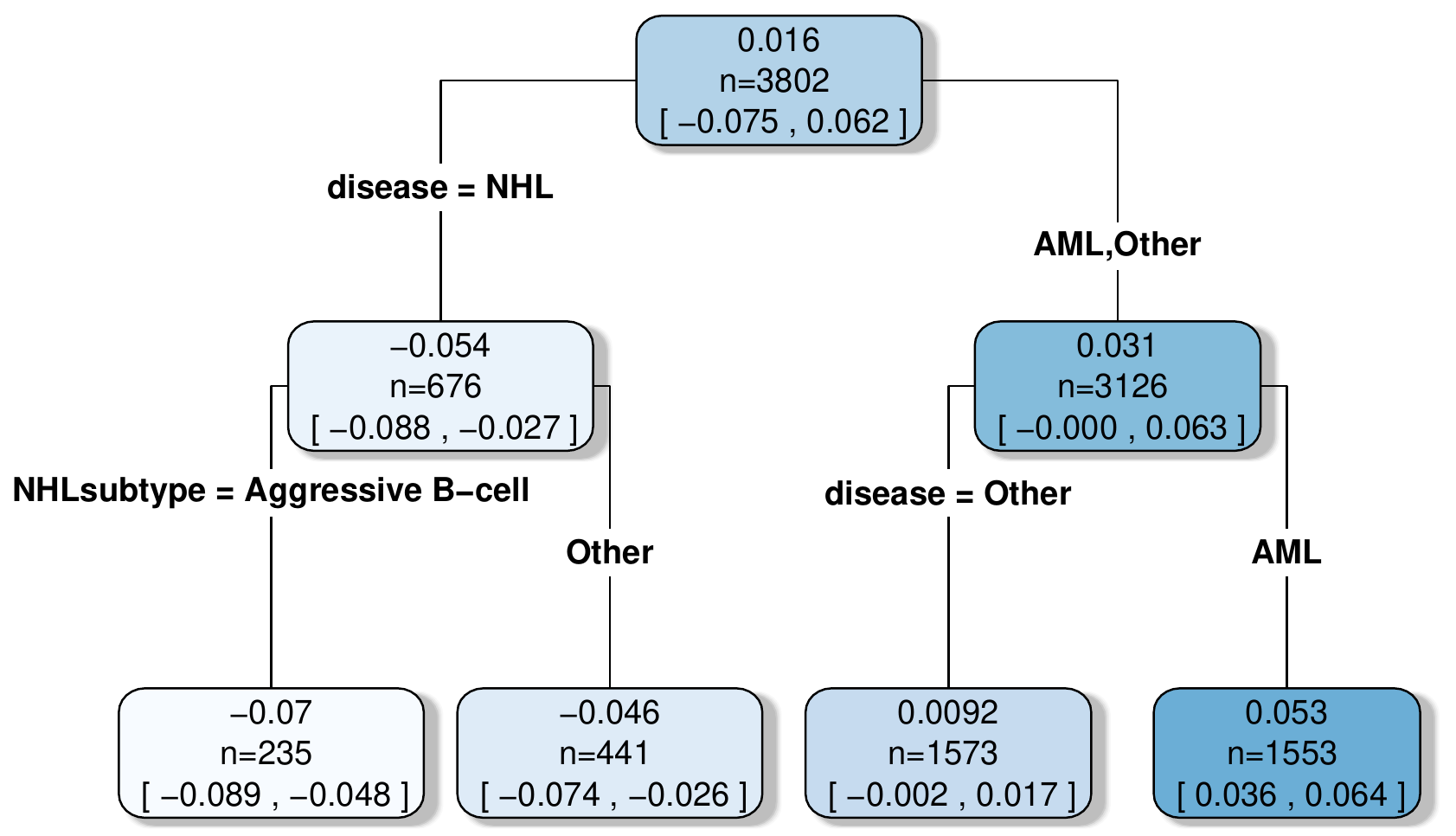}
\end{center}
\caption{Tree fit to the posterior mean treatment differences.  Values in each node represent the posterior mean and 95\% credible intervals for the average treatment effect of the subgroup of patients represented in that node.}
\label{fig:treefit}
\end{figure}

As pointed out earlier, one drawback of the BART based ITR is that it does not lead to a simple interpretable rule.  Next we apply the fit the fit technique to approximate the BART based ITR with an interpretable treatment rule which has nearly as good performance.  In order to do this, the posterior mean treatment differences in 1 year survival are treated as the ``data'', and we try and fit an interpretable single tree to this data.  In order to do this, a sequence of trees are fit, where variables are added sequentially to the set of potential splitting variables of the tree in a stepwise manner to improve the fit.  Once the change in $R^2$ is less than 1\% with addition of a new variable, the procedure ends and the current tree is used as the approximation to the BART ITR. The results of the final tree fit are shown in Figure~\ref{fig:treefit}; $R^2$ between the tree fit and the posterior mean treatment differences is 97\%.  The first splitting variable used, Non-Hodgkins Lymphoma (NHL) disease vs. other disease, was sufficient to match the BART ITR exactly in terms of selection of conditioning regimen; Patients with NHL have approximately 5\% better 1 year OS with Flu/Bu conditioning, while patients without NHL have approximately 3\% better 1 year OS with Flu/Mel conditioning.  A second level of splitting variables provided further resolution on the magnitude of the treatment benefit for disease subgroups, but did not affect the directional benefit.  Patients with AML have approximately 5\% better 1 year OS with Flu/Mel conditioning, while patients with other diseases have approximately 1\% better 1 year OS with Flu/Mel conditioning.  Similarly, among patients with NHL disease, those with aggressive B-cell subtype may experience slightly more benefit from Flu/Bu conditioning compared to those with other subtypes (7\% vs. 5\%).  Note that in this case, this simple rule completely matched the BART based ITR, and so the value function associated with this rule also matches the BART ITR value function in the previous figure.  This type of approximation to a BART based ITR can greatly facilitate communication with clinicians on how the ITR works and what types of patients benefit from one treatment vs. another. 

\section{Discussion}
\label{Discussion}
In this article, we presented a framework for identifying optimal
individualized treatment rules using Bayesian Additive Regression
Trees.  BART is a flexible fully nonparametric prediction model which
can handle complex functional forms as well as interactions among
variables, and therefore it is well suited for examining treatment
interactions which drive ITRs.  There are two main advantages to using BART to identify an ITR.  First, our proposed method has excellent
performance when benchmarked against existing methods, including other
flexible prediction methods as well as policy search methods such as
outcome weighted learning; overall it performed better than or comparable to other existing methods across a range of ITR simulation settings established in other papers.  Second, our method provides direct inference on the value of the BART ITR as well as the Optimal ITR.  This requires incorporation of both the uncertainty in the prediction model, as well as uncertainty in the individual patient treatment selection that depends on the prediction model.  Both can be handled in a straightforward manner using the posterior samples for the prediction model function in the Bayesian framework.  In contrast, it is not clear how to do this for policy search methods which do not provide a direct prediction model for outcome, as well as for other flexible models such as random forests which do not directly provide uncertainty measures.    

We observed that the BART model tends to shrink the treatment effect
estimates for smaller sample size leading to underestimation of the
true value of the Optimal ITR. Larger sample sizes do overcome this
shrinkage from the prior.  Further consideration of ways to reduce
this unidirectional bias could provide better inference on the value
function for small sample sizes. 

One limitation of our proposed method is that the BART method generates a ``black box'' prediction model which is difficult to explain to clinicians. However, we showed how the posterior mean differences available from the model can be fed into a tree procedure to yield an interpretable ITR which provides a close approximation to the BART ITR and which can be readily explained to clinicians.  A recent article \cite{SivaMuel17} describes the use of BART and utility specifications with the goal of identifying subgroups with elevated treatment effects, in contrast to finding an individualized treatment rule.  

Our proposed method uses ``off-the-shelf'' BART software; minimal
processing is needed to obtain posterior inference under each
treatment condition for patients in a test dataset. Inference for
the value of an ITR is also readily available. While
BART can be computationally demanding as a Markov Chain
Monte Carlo (MCMC) technique, it can be parallelized to save
computational time since the  chains do not share information beyond
the data itself.

\section*{Acknowledgements}

Work of all authors is supported in part by the Advancing a Healthier Wisconsin Endowment at the Medical College of Wisconsin.  The authors would like to thank Dr. Menggang Yu from the University of Wisconsin for providing software to implement the ROWSI method in the simulation study.

 \newpage

\bibliographystyle{abbrv}
\bibliography{bibrefsmmr}

\end{document}